# Breakdown of agreement between non-relativistic and relativistic quantum dynamical predictions in the non-relativistic regime

Boon Leong Lan[*], Mehdi Pourzand and Rui Jian Chu

*Electrical and Computer Systems Engineering & Advanced Engineering Platform, School of Engineering, Monash University, 47500 Bandar Sunway, Malaysi*a

[*]lan.boon.leong@monash.edu

**To study quantum dynamics in the non-relativistic regime, the standard practice is to use non-relativistic quantum mechanics, instead of the relativistic theory, because it is thought the approximate non-relativistic result is always close to the relativistic one. Here we present a theoretical argument that this expectation is not true in general. In addition, supporting numerical evidence for the free rotor and hydrogen atom also shows the agreement between the two theories can break down quickly. For the radial Rydberg wave packet in hydrogen atom, the breakdown can occur before spontaneous emission and thus could be tested experimentally. Our surprising result shows relativistic quantum mechanics must be used, instead of the approximate non-relativistic theory, to correctly study quantum dynamics in the non-relativistic regime after the breakdown time. This paradigm shift opens a new avenue of research in a wide range of fields from atomic to molecular, chemical and condensed-matter physics.**

**Introduction**

Relativistic quantum mechanics is an essential fundamental theory in many fields of physics, ranging from atomic to molecular, chemical and condensed matter physics [1]. Application of the theory to study quantum dynamics requires the wave function solution to the time-dependent Dirac equation. However, since few analytical solutions exist for problems of physical interest, numerical solutions [2,3] must be obtained but the calculations are typically computationally challenging. Alternatively, the time-dependent probability density constructed from an ensemble of relativistic classical trajectories is used to approximate the relativistic quantum probability density [4-7].

In the non-relativistic regime, the standard practice is to use the non-relativistic time-dependent quantum wave function to approximate the relativistic wave function. However, it is not known whether the non-relativistic time-dependent wave function and quantities derived from it are always close approximation to the relativistic counterparts as conventionally expected [8]. This important fundamental question in quantum physics has not been addressed so far – the accuracy of the non-relativistic approximation to relativistic dynamics in the non-relativistic regime has only been studied in the classical context [9-15]. Here we present a theoretical argument that the non-relativistic quantum approximation is, generally, not always close to the relativistic quantum dynamics in the non-relativistic regime, and provide supporting numerical evidence for the free rotor and hydrogen atom.

**Results and discussion**

The relativistic time-independent Dirac equation for an electron, in which the rest mass energy is subtracted from the Dirac operator $\widehat{H}_D$, is given by [16,17]

$$(\widehat{H}_D - m_0 c^2 I_4) \begin{bmatrix} \chi_n \\ \eta_n \end{bmatrix} = E_n \begin{bmatrix} \chi_n \\ \eta_n \end{bmatrix} \quad (1)$$

where $\chi_n$ and $\eta_n$ are two-component spinors. The equation has positive and negative energy solutions, where the former describes electron and the latter describes positron [16]. Here, we focus on the typical situations in atomic to condensed matter physics where the energies involved are too small to trigger electron-positron pair-creation processes [16,18,19]. (If pair-creation processes occur, QED would have to be used instead [16,18].) In such cases, an initial superposition of positive-energy eigenstates, which describes the electron, remains a superposition of positive-energy eigenstates – therefore, only the positive-energy subspace needs to be considered for the relativistic dynamics of the electron [16].

In the non-relativistic regime [17], the upper component of the relativistic energy eigenstate is much larger than the lower component for positive energy solutions, i.e.

$$\chi_n \gg \eta_n. \quad (2)$$

Furthermore [16,17], $\chi_n$ is close to the non-relativistic energy eigenstate $\varphi_n$ (which is also a two-component spinor)

$$\chi_n \approx \varphi_n \tag{3}$$

and the relativistic energy $E_n$ is close to the non-relativistic energy $\varepsilon_n$

$$E_n \approx \varepsilon_n, \tag{4}$$

where $\varphi_n$ and $\varepsilon_n$ are, respectively, the eigenfunctions and eigenvalues of the non-relativistic Pauli operator $\widehat{H}_{NR}$

$$\widehat{H}_{NR}\varphi_n = \varepsilon_n \varphi_n. \tag{5}$$

The relativistic and non-relativistic time-dependent wave functions are, respectively, given by

$$\psi_R(t) = \sum A_n(0) e^{-iE_n t/\hbar} \chi_n \tag{6}$$

$$\psi_{NR}(t) = \sum a_n(0) e^{-i\varepsilon_n t/\hbar} \varphi_n = \sum a_n(0) e^{-iE_n t/\hbar} e^{i\delta_n t/\hbar} \varphi_n \tag{7}$$

where $\delta_n = E_n - \varepsilon_n$ is the small difference between the relativistic energy $E_n$ and non-relativistic energy $\varepsilon_n$. For the same relativistic and non-relativistic expansion coefficients, i.e., $A_n(0) = a_n(0)$, the relativistic and non-relativistic initial wave functions are close since the relativistic and non-relativistic states, $\chi_n$ and $\varphi_n$, are close. However, it is evident from Eqs. (6) and (7) that the relativistic and non-relativistic wave functions will not always be close to each other because of the extra time-dependent phase-factor $e^{i\delta_n t/\hbar}$, where the phase grows linearly with time, in each term of the sum in Eq. (7) for the non-relativistic wave function. This implies that the relativistic and non-relativistic probability densities and expectation values will, likewise, not always be close. This is also the case for the relativistic and non-relativistic autocorrelation functions, where the autocorrelation function of the wave function at time $t$ with the initial wave function is defined as

$$C(t) \equiv \langle \psi(t) | \psi(0) \rangle. \tag{8}$$

The relativistic and non-relativistic autocorrelation functions can be, respectively, expressed as

$$C_R(t) = \sum |A_n(0)|^2 e^{iE_n t/\hbar} \tag{9}$$

$$C_{NR}(t) = \sum |a_n(0)|^2 e^{iE_n t/\hbar} e^{-i\delta_n t/\hbar}. \tag{10}$$

In the non-relativistic regime, it is also evident from the expressions above that the two autocorrelation functions will not always be close to each other, for the same expansion coefficients, because of the extra time-dependent phase-factor $e^{-i\delta_n t/\hbar}$ in Eq. (10) for the non-relativistic autocorrelation function, where $\delta_n$ is the small difference between the relativistic and non-relativistic energies.

For the free rotor, i.e., a free electron constrained to move in a circle of radius $R$, both the relativistic and non-relativistic initial expansion coefficients are chosen to be

$$A_n(0) = a_n(0) = \left(\frac{2\sigma_0^2}{\pi}\right)^{1/4} exp(-in\theta_0) \; exp(-\sigma_0^2(n-\bar{n})^2), \tag{11}$$

where $|A_n(0)|^2 = |a_n(0)|^2$ is a narrow Gaussian centered at $\bar{n}$, with variance $1/\sigma_0^2$. The non-relativistic regime requires (see Methods) that the principal quantum number $n$ of these states in the superposition satisfy $|n\hbar| \ll m_0 cR$. For $\sigma_0 < 1$, the relativistic and non-relativistic initial wave functions are approximately Gaussians in the interval $[0, 2\pi]$ with means $\theta_0$ and $\bar{n}\hbar$, variances $\sigma_0^2$ and $\hbar^2/4\sigma_0^2$, for, respectively, the angle and angular momentum [20]. In both theories, the mean and variance of the angular position change with time (see Methods) but the mean and variance of the angular momentum do not. Since the relativistic and non-relativistic values are initially the same for the mean and also variance of the angular momentum, they remain the same. Here we present a representative example to illustrate the typical result of comparing the relativistic and non-relativistic expectation values and probability densities for the angular position of the free rotor in the non-relativistic regime. Hartree atomic units are used in the numerical calculations, where $m_0 = \hbar = 1, c = 137.035999037$ [21]. Fig. 1 shows the relative difference between the relativistic and non-relativistic values fluctuates as it grows, for the mean and variance of the angular position. The maximum magnitude of the relative difference for the means and variances are 17% and 51%, 36% and 193%, 38% and 234% in a small time interval $4\pi \times 10^6$ $a.u.$ $(3.0 \times 10^{-10}$ $s)$ centered at, respectively, time $t_1 \approx 2.2 \times 10^{14}$ $a.u.$ $(0.0053$ $s)$, $t_2 \approx 2.1 \times 10^{15}$ $a.u.$ $(0.051$ $s)$, $t_3 \approx 3.1 \times 10^{16}$ $a.u.$ $(0.75$ $s)$ indicated in the figure. Fig. 2 shows the relativistic and non-relativistic probability densities are, correspondingly, increasingly dissimilar from one time interval to the next, where the latter bears no resemblance to the former in the last time interval centered at $t_3$. The non-relativistic wave packet for the free rotor undergoes perfect (i.e., exact) full revival [22,23], which is seen in Fig. 2(a) to 2(c) since the time interval $4\pi \times 10^6$ $a.u.$ is the non-relativistic revival time for the $\bar{n}$ in this example. In the relativistic case, for times larger than the revival time (which is very close to the non-relativistic revival time) but much smaller than the super-revival time $(2.4 \times 10^{17}$ $a.u.)$, the wave packet also undergoes full revivals but the revivals are not perfect (Figs. 2(d) and 2(e)). However, at times appreciable compared to the super-revival time, the full-revival sequence collapses (Fig. 2(f)). The increasingly different relativistic and non-relativistic results in Figs. 1 and 2 are essentially due to the extra phase-factor $e^{-i\delta_n t/\hbar}$ in each component of the relativistic wave function (Eq. (6)) compared to the corresponding component of the non-relativistic wave function (Eq. (7)), where the extra phase $-\delta_n t/\hbar$, which grows with time, is different for different component. This also explains why the relativistic wave packet does not revive perfectly like the non-relativistic wave packet.

For hydrogen atom, the relative difference between the relativistic and non-relativistic energies (the formula for the energies are given in Methods) increases from $n = 1$ to $n = 2$, but thereafter it decreases with $n$ – see Fig. 3(a). Since the maximum relative difference is only 0.0017%, the relativistic and non-relativistic energies are very close for all $n$. Here, we focus on a superposition of states, with orbital angular momentum $l = 1$ and total angular momentum $j = 1/2$ (and, therefore $m_l = 0$), centered on a high principal quantum number $\bar{n}$, i.e., a radial Rydberg wave packet [24]. Furthermore, we focus on the autocorrelation function because it is related to the ionization signal in a pump-probe experiment [22, 24-27]. The squared-amplitude of the initial expansion coefficients in both theories are chosen to be a narrow Gaussian, with variance $1/\sigma_0^2$, centered at the high $\bar{n}$. The non-relativistic autocorrelation function, Eq. (10), can be re-written as

$$C_{NR}(t) = \sum |A_n(0)|^2 \exp\left[iE_n\left(t - \frac{\delta_n}{E_n}t\right)/\hbar\right] \quad (12)$$

where $\delta_n = E_n - \varepsilon_n$ is the small difference between the relativistic energy $E_n$ and non-relativistic energy $\varepsilon_n$. Since $|A_n(0)|^2$ is a narrow Gaussian centered at a high $\bar{n}$, the relative energy difference $\frac{\delta_n}{E_n}$ is essentially the same, approximately $\frac{\delta_{\bar{n}}}{E_{\bar{n}}}$, for different $n$ in the narrow range centered at $\bar{n}$ – see Fig. 3(a). Thus, Eq. (12) implies

$$C_{NR}(t) \approx C_R(t') \quad (13)$$

where $t' = t - \frac{\delta_{\bar{n}}}{E_{\bar{n}}}t$. Over a small time interval, $\frac{\delta_{\bar{n}}}{E_{\bar{n}}}t$ is approximately constant, therefore the non-relativistic autocorrelation function is approximately the relativistic autocorrelation function that is time shifted by this constant – Figs. 3 illustrates this time shift for two values of $\bar{n}$: 40 (Figs. 3(b) and 3(c)) and 300 (Figs. 3(d) to 3(f)), where Hartree atomic units are used in the numerical calculations. The maximum time in the plot is $10T_{sup} \approx 80 \text{ ns}$ and $100T_{sup} \approx 800 \text{ ns}$, respectively, in Figs. 3(b) and 3(c), and $10T_{sup} \approx 2 \text{ ms}$, $40T_{sup} \approx 7 \text{ ms}$ and $61T_{sup} \approx 11 \text{ ms}$, respectively, in Figs. 3(d) to 3(f), where $T_{sup}$ is the relativistic super-revival time (which is defined similarly as the non-relativistic counterpart [22,23]) for the corresponding $\bar{n}$. The shape of each peak of the non-relativistic squared-amplitude autocorrelation function is close to the corresponding relativistic one, but the non-relativistic peak occurs at a later time compared to its relativistic counterpart. The time difference between the non-relativistic and relativistic peaks in each pair is 0.15 ps and 1.5 ps, respectively, in Figs. 3(b) and 3(c), and 0.5 ns, 2 ns and 3 ns, respectively, in Figs. 3(d) to 3(f). These time differences are close to the estimated time shift $\frac{\delta_{\bar{n}}}{E_{\bar{n}}}t$, which increases with time for fixed $\bar{n}$. For $\bar{n} = 300$, the relativistic and non-relativistic peaks in each pair are time shifted dramatically in Fig. 3(f) – they no longer overlap, and the non-relativistic peak occurs only after another relativistic peak has occurred after the relativistic counterpart. The relativistic super-revival time is several (i.e., more than two) orders of magnitude less than the spontaneous lifetime of the Rydberg state [23,24]. It is therefore possible to test the different non-relativistic and relativistic autocorrelation functions for the radial Rydberg wave packet experimentally. One possibility is to use the optical Ramsey pump-probe method [24-27], which uses two identical laser pulses with a time delay. For a fixed delay time, the total Rydberg population changes with the phase of the probe pulse [26]. The amplitude of the autocorrelation function at the delay time is measured by measuring the rms value of the phase-dependent Rydberg population using field ionization, which has a detection efficiency of almost 100% [26].

In general, if the initial expansion coefficients are strongly centered about $\bar{n}$, the non-relativistic quantum dynamics is close to the relativistic quantum dynamics in the non-relativistic regime for, approximately, time $t \ll T_{critical}$, where

$$T_{critical} = \hbar/|\delta_{\bar{n}}| \quad (14)$$

and $\delta_{\bar{n}}$ is the small difference between the relativistic and non-relativistic energies of the state $\bar{n}$. This time estimate, which is inversely proportional to $\delta_{\bar{n}}$, is obtained by requiring the extra phase $\delta_{\bar{n}} t/\hbar$ for the dominant term in the non-relativistic Eqs. (7) and (10) to be much less than one radian. For the hydrogen atom example, $T_{critical}$ is, respectively, 59 ns and 0.02 ms for $\bar{n} = 40$ and $\bar{n} = 300$, and for the free rotor example, $T_{critical}$ is $1.5 \times 10^{17}$ $a.u.$ for $\bar{n} = 1$, which are consistent with our numerical results.

**Conclusion**

Our unexpected result shows relativistic quantum mechanics must be used, instead of the approximate non-relativistic theory, to correctly study quantum dynamics in the non-relativistic regime after the breakdown of the non-relativistic approximation. The breakdown time can be estimated using Eq. (14). However, how long it actually takes for the break down to occur would have to be specifically studied numerically for each system. Such studies in the future should, for example, include coherent control scenarios [28-30], where wave packets are steered by light. Understanding the accuracy of the non-relativistic laser-driven quantum dynamics relative to the relativistic theory in this context is crucial for the experimental realizations of the coherent control strategies.

**Methods**

For a free electron constrained to move in a circle of radius $R$, the Dirac operator is, following Strange [31], given by

$$\widehat{H}_D = \frac{c}{R} \alpha \hat{p}_\theta + \beta m_0 c^2 \qquad (15)$$

where $\hat{p}_\theta = -i\hbar \frac{\partial}{\partial \theta}$ is the angular momentum operator,

$$\alpha = \begin{bmatrix} 0_2 & \sigma_2 \\ \sigma_2 & 0_2 \end{bmatrix}, \qquad \beta = \begin{bmatrix} I_2 & 0_2 \\ 0_2 & -I_2 \end{bmatrix}, \qquad \sigma_2 = \begin{bmatrix} 0 & -i \\ i & 0 \end{bmatrix},$$

and $0_2$ and $I_2$ are, respectively, zero and identity $2 \times 2$ matrices. Eq. (1) can be solved exactly to yield the relativistic energy

$$E_n = m_0 c^2 \sqrt{1 + \frac{\hbar^2 n^2}{m_0^2 c^2 R^2}} - m_0 c^2, \qquad n = 0, \pm 1, \pm 2, \ldots \qquad (16)$$

and energy eigenstate $\begin{bmatrix} \chi_n \\ \eta_n \end{bmatrix}$, where

$$\chi_n = N_n e^{in\theta} \begin{bmatrix} 1 \\ 0 \end{bmatrix} \qquad (17)$$

$$\eta_n = N_n e^{in\theta} \left[\begin{pmatrix} 0 \\ \frac{ic\hbar n/R}{E_n + 2m_0 c^2} \end{pmatrix}\right], \tag{18}$$

and the normalization constant is given by

$$N_n = \frac{1}{\sqrt{2\pi \left[1 + \frac{c^2 \hbar^2 n^2}{R^2 (E_n + 2m_0 c^2)^2}\right]}}. \tag{19}$$

(We note that if the rest mass energy is not subtracted from the Dirac operator in Eq. (1), the relativistic energy $E_n$ does not have the $-m_0 c^2$ term, and $\eta_n$ and $N_n$ are the same except the $2m_0 c^2$ term is replaced by $m_0 c^2$, as given in [31].) The relativistic energy eigenstate is also the eigenstate of the angular momentum operator, with eigenvalue $n\hbar$. In the non-relativistic regime, where the angular momentum is small, i.e., $|n\hbar| \ll m_0 cR$ [31], the relativistic energy is approximately the non-relativistic energy

$$E_n \approx \frac{\hbar^2 n^2}{2m_0 R^2} = \varepsilon_n \tag{20}$$

and the $\chi_n$ component of the relativistic energy eigenstate, which is $\gg \eta_n$, is approximately the non-relativistic energy eigenstate $\varphi_n$

$$\chi_n \approx \frac{1}{\sqrt{2\pi}} e^{in\theta} \begin{bmatrix} 1 \\ 0 \end{bmatrix} = \varphi_n \tag{21}$$

since $N_n \approx \frac{1}{\sqrt{2\pi}}$.

For the free rotor, the relativistic mean and variance of the angular position at time $t$ are, respectively, given by

$$\langle\theta\rangle_t = \pi + \sum_{r \neq s} \Omega_{r,s} \frac{A_r^*(t) A_s(t)}{i(s-r)} \tag{22}$$

$$\langle\theta^2\rangle_t - \langle\theta\rangle_t^2 = \frac{4\pi^2}{3} - 2 \sum_{r \neq s} \Omega_{r,s} A_r^*(t) A_s(t) \left[\frac{\pi i}{(s-r)} - (s-r)^{-2}\right] - \langle\theta\rangle_t^2 \tag{23}$$

where

$$\Omega_{r,s} = 2\pi N_r N_s + \sqrt{(1 - 2\pi N_r^2)(1 - 2\pi N_s^2)}. \tag{24}$$

The non-relativistic mean and variance of the angular position are given by the same formulae [32], except $\Omega_{r,s}$ does not appear in the sums and $A_n(0)e^{-iE_nt/\hbar}$ is replaced by $A_n(0)e^{-i\varepsilon_n t/\hbar}$ for the expansion coefficient $A_n(t)$.

For hydrogen atom, the relativistic energy, with the rest mass energy subtracted, is given by [33]

$$E_{n,j} = m_0 c^2 \left\{ 1 + \frac{\alpha^2}{\left[n - j - 1/2 + \sqrt{(j + 1/2)^2 - \alpha^2}\right]^2} \right\}^{-\frac{1}{2}} - m_0 c^2 \qquad (25)$$

where $n = 1, 2, \ldots$ is the principal quantum number, $j$ is the total angular momentum quantum number ($j = \left|l - \frac{1}{2}\right|, l + \frac{1}{2}$ where $l = 0, 1, \ldots, n - 1$ is the orbital angular momentum quantum number) and $\alpha$ is the fine structure constant. The relativistic energy can be expanded as [33]

$$E_{n,j} = -\frac{m_0 c^2 \alpha^2}{2}\left[\frac{1}{n^2} + \frac{\alpha^2}{n^4}\left(\frac{n}{j + \frac{1}{2}} - \frac{3}{4}\right)\right] + \cdots \qquad (26)$$

where the first term is the non-relativistic energy, which depends only on $n$.

**Acknowledgement**

This work was supported by a Fundamental Research Grant FRGS/1/2013/ST02/MUSM/02/1.

**Figure Captions**

**Figure 1. Comparison of the relativistic and non-relativistic expectation values for the angular position of the free rotor.** In this example, the parameters for the initial expansion coefficients in both theories are $\sigma_0 = 0.271$, $\theta_0 = \pi$, $\bar{n} = 1$. The radius $R$ of the electron's circular path is 1000 a.u., thus, the mean angular momentum is small, i.e., $\bar{n}\hbar \ll m_0 cR$. **(a), (b)** Relative difference between the means. **(c), (d)** Relative difference between the variances. In all cases, the relative difference is calculated as (non-relativistic value - relativistic value)/relativistic value.

**Figure 2. Comparison of the relativistic and non-relativistic probability densities for the angular position of the free rotor.** The comparison is for the example in Fig. 1. **(a), (b), (c)** Non-relativistic probability density in a small time interval of $2T = 4\pi \times 10^6\ a.u.$ centered at, respectively, time $t_1 \approx 2.2 \times 10^{14}\ a.u.$, $t_2 \approx 2.1 \times 10^{15}\ a.u.$ and $t_3 \approx 3.1 \times 10^{16}\ a.u.$ indicated in Fig. 1. **(d), (e), (f)** The corresponding relativistic probability densities.

**Figure 3. Comparison of the relativistic and non-relativistic energies and autocorrelation functions for hydrogen atom ($l = 1, j = 1/2$).** **(a)** Relative energy difference [(relativistic energy – non-relativistic energy)/relativistic energy] versus principal quantum number. **(b), (c), (d), (e), (f)** Squared-amplitude autocorrelation functions: relativistic (blue), non-relativistic (red), calculated using the same narrow Gaussian, with standard deviation $1/\sigma_0$ where $\sigma_0 = 0.505$, centered at $\bar{n}$ for the squared-amplitude of the initial expansion coefficients. In (b) to (d), the adjacent relativistic and non-relativistic peaks constitute a pair. In (e) and (f), the pair of arrows indicates a pair of relativistic and non-relativistic peaks.

**Figure 1**

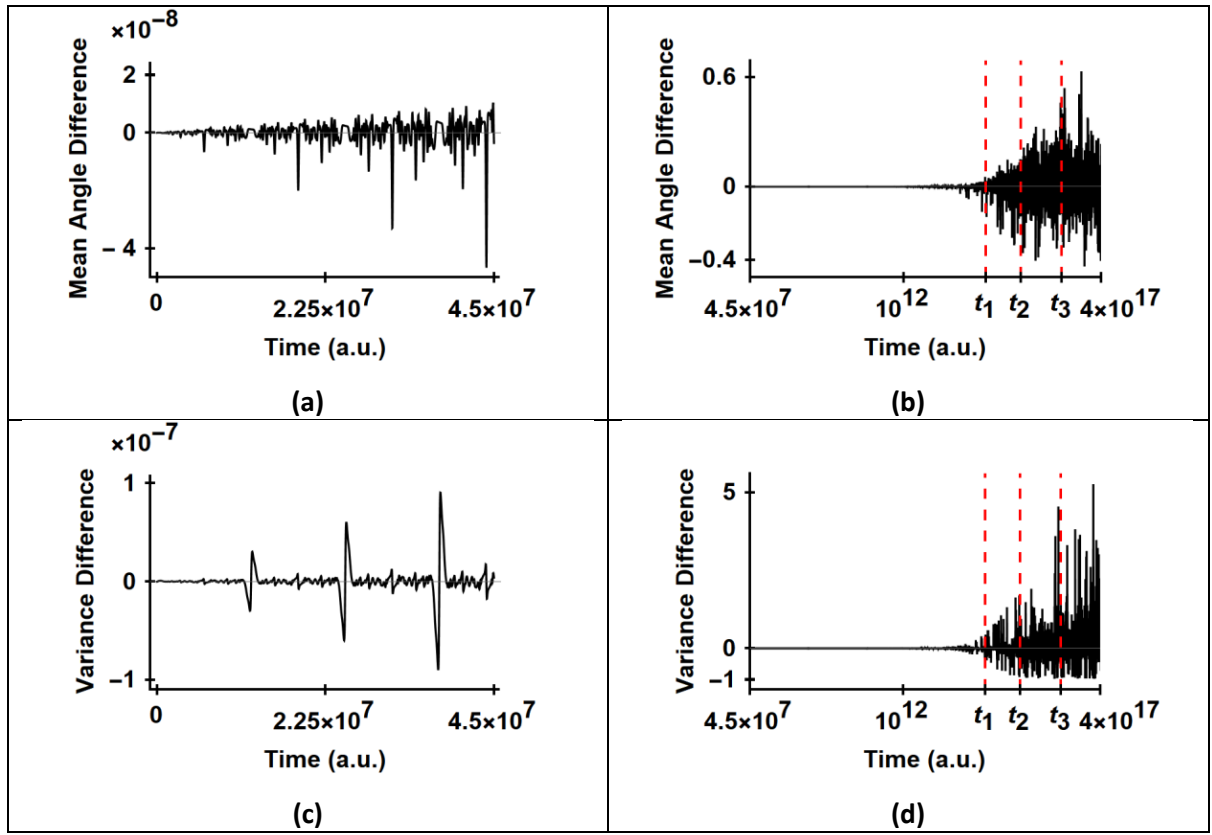

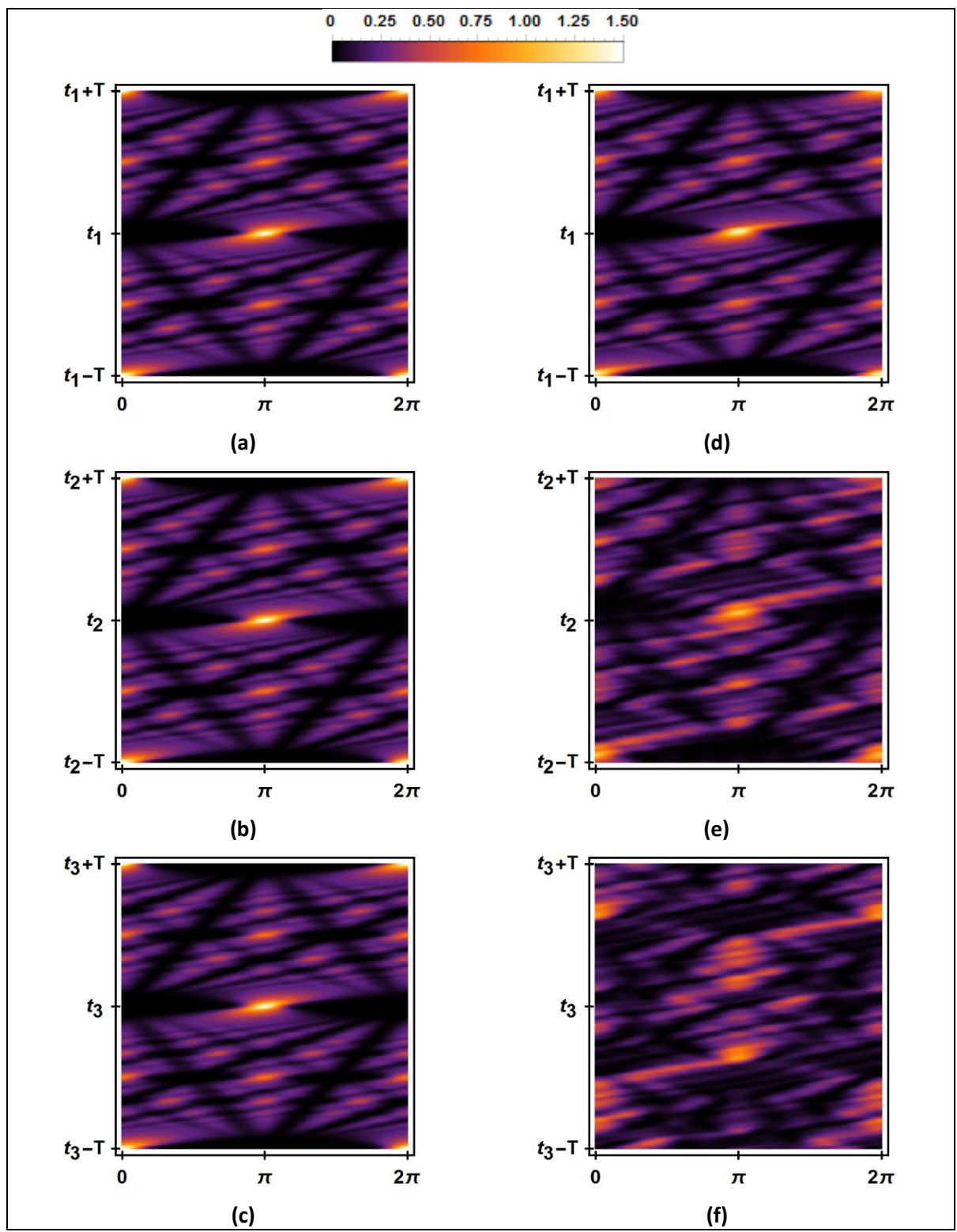

**Figure 3**

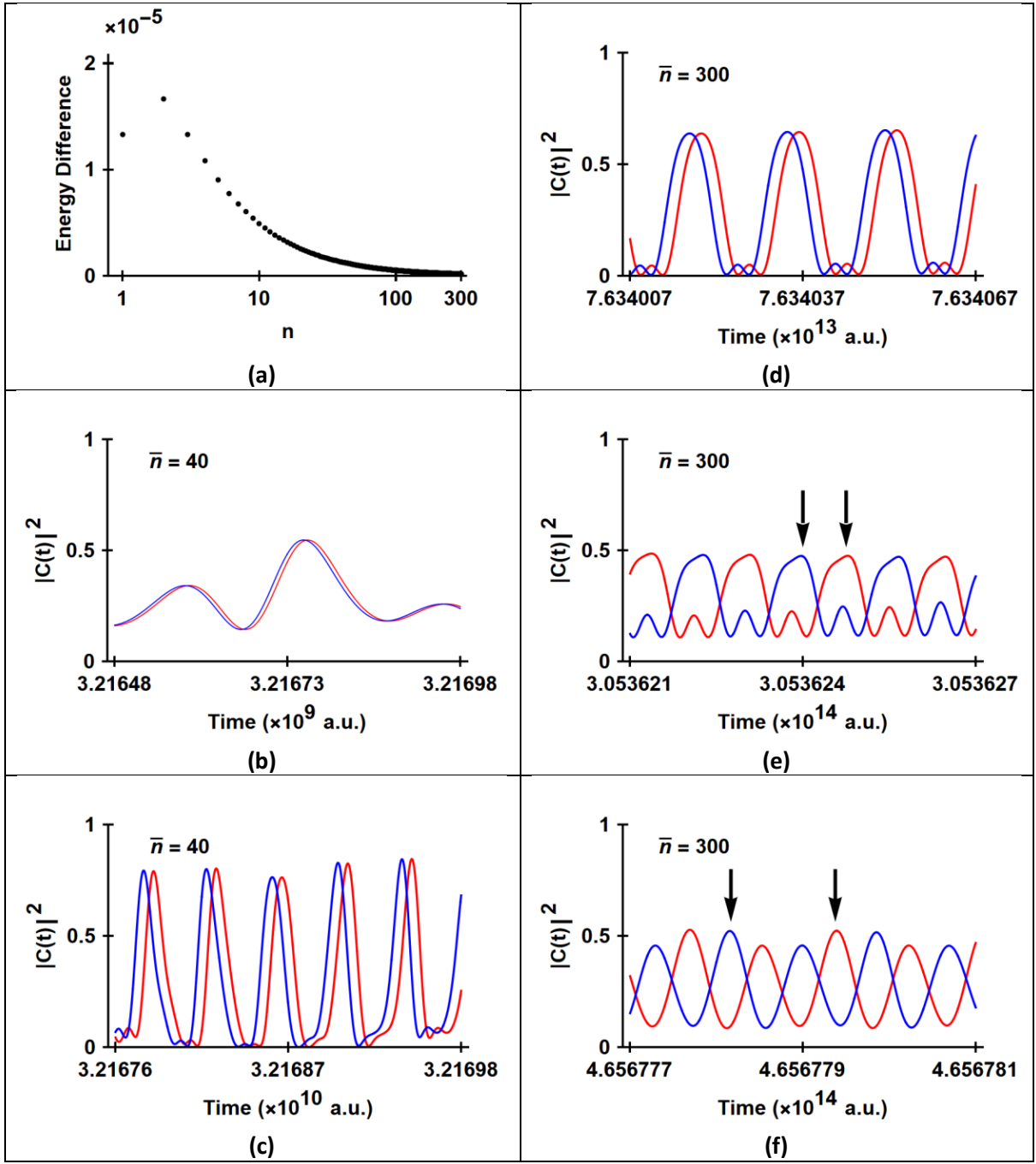